\documentclass[12pt]{article}

 \usepackage{amssymb}
 \usepackage{amsmath}
 \usepackage{amsfonts}
 \usepackage{color}
 \newcommand {\be}{\begin{equation}}
 \newcommand {\ee}{\end{equation}}
 \newcommand {\bea}{\begin{array}}
 
 \newcommand {\eea}{\end{array}}
 \newcommand {\pa}{\partial}
 \newcommand {\al}{\alpha}
 \newcommand {\de}{\delta}
 \newcommand {\ta}{\tau}
 \newcommand {\ga}{\gamma}
 \newcommand {\ep}{\epsilon}
 \newcommand {\si}{\sigma}

\begin{document}

  \thispagestyle{empty}

  \vspace{2cm}

  \begin{center}
    \font\titlerm=cmr10 scaled\magstep4
    \font\titlei=cmmi10 scaled\magstep4
    \font\titleis=cmmi7 scaled\magstep4
  {\bf  Constraint structure and Hamiltonian treatment of
      Nappi-Witten model}

    \vspace{1.5cm}
    \noindent{{
        M. Dehghani\footnote{mdehghani@ph.iut.ac.ir}
        A. Shirzad\footnote{shirzad@ipm.ir}
         }}\\
    \vspace{0.8cm}

   {\it Department of Physics, Isfahan University of Technology \\
P.O.Box 84156-83111, Isfahan, IRAN, \\
School of Physics, Institute for Research in Fundamental Sciences (IPM)\\
P.O.Box 19395-5531, Tehran, IRAN.}

  \end{center}

  \vskip 2em

  \begin{abstract}
We investigate the Hamiltonian analysis of Nappi-Witten model (WZW
action based on non semi simple gauge group) and find a time
dependent non-commutativity by canonical quantization. Our
procedure is based on constraint analysis of the model in two
parts. A first class analysis is used for gauge fixing the
original model following by a second class analysis in which the
boundary condition are treated as Dirac constraints. We find the
reduced phase space by imposing our second class constraints on
the variables in an extended Fourier space.
  \end{abstract}




 \section{Introduction \label{sec1}}
Treating boundary conditions as Dirac constrains  has been
considered in the recent years by so many authors \cite{ref0,
refsh0, Tlee1, baner1}. This approach has been used first in
studying the Polyakov string coupled to a B-field. The common
feature of all works is non commutativity of the coordinate fields
on the boundaries which may lie on some brains, as first predicted
by \cite{SW}. However, there are different approaches in defining
the constraints and investigating their consistency in time. We
have reviewed the whole subject in our previous work \cite{ref3}
and showed if we impose the set of constraints on the Fourier
expansions of the fields, the redundant modes will be omitted in a
natural way.

For simple physical models obeying linear equations of motion, the
ordinary Fourier expansion gives appropriate coordinates to reach
the reduced phase space. In other words, the infinite set of
second class constraints emerging as the result of boundary
conditions, forces us to omit a number of Fourier modes. However,
ordinary Fourier transformation is not essential for quantization;
it is just one tool that acts well for most physical models at
hand. In the general case one should search for "appropriate
coordinates", in which imposing the set of second class
constraints is equivalent to omitting some canonical pairs from
the theory.

In this paper we study the constraint structure of the
Nappi-Witten model in the Hamiltonian formalism. This model
acquires complicated boundary conditions so that the ordinary
Fourier expansion seems inadequate to impose the whole set of
constraints which emerge from the boundary conditions.
Nevertheless, the Nappi-Witten model, on its own grands, is an
attractive one since it describes a non semi simple gauge group as
well as giving time dependent non commutativity in some gauges
\cite{ref2}. Our next interest is to emphasize that solving the
equations of motion is not necessarily needed for quantizing a
theory; the only necessity is finding the dynamics of the
constraints and construct their algebra with the Hamiltonian such
that they remain consistent with time on the constraint surface.

We give a precise Hamiltonian treatment of the model in which the
constraint structure is followed step by step from the initial
action to the final reduced phase space. In section 2 we introduce
the model and find primary and secondary constraints of the
system. Section 3 is devoted to fixing the gauge by introducing
appropriate gauge fixing conditions. In section 4 we follow our
strategy of treating the boundary conditions as primary Dirac
constraints and follow their consistencies. The boundary
conditions which come from the original action, in fact, make the
system more complicated. So, it is not possible to write down the
solutions in a closed form similar to a simple Fourier expansion
(see reference \cite{ref1}). We try to find a basis which is
appropriate for imposing the infinite set of constraints in
section 5. In section 6 we will give our concluding remarks and
will compare our results with parallel approaches.

\section{Hamiltonian structure of the model \label{sec2}}
The Nappi-Witten model describes a 4-component bosonic string
$X_a=(a_1,a_2,u,v)$ living in the background metric $G_{ab}(X)$
and coupled to a $B$-field. The action is given as:
 \be S=\int d^2\si\bigg[\sqrt{-g}g^{ij}G_{ab}\pa_iX^a\pa_jX^b+B_{ab}\ep^{ij}\pa_iX^a\pa_jX^b\bigg], \label{f1}\ee
 where
\be G(X)=\left(\begin{array}{llll}
1 & 0 & \frac{a_2}{2} & 0 \\
0 & 1 & -\frac{a_1}{2} & 0 \\
\frac{a_2}{2} & -\frac{a_1}{2} & b & 1 \\
0 & 0 & 1 & 0
\end{array}\right)
,B(X)=\left(\begin{array}{llll}
0 & u & 0 & 0 \\
-u & 0 & 0 & 0 \\
0 & 0 & 0 & 0 \\
0 & 0 & 0 & 0
\end{array}\right). \label{f2}\ee
The special form of $G(X)$ and $B(X)$ are chosen so that the gauge
group of the model is non semi-simple \cite{ref1}. The metric
field can be written in terms of the following variables: \be
\begin{array}{lll} N_1=\frac{1}{g^{00}\sqrt{-g}}, &
N_2=-\frac{g^{01}}{g^{00}}, & N_3=\sqrt{-g}=
\frac{1}{\sqrt{(g^{01})^2-g^{00}g^{11}}}.
\\  \end{array} \label{f3}\ee
In terms of the variables $X_a$ and $N_\alpha$ the action becomes:
\be S=\int
d^2\si\bigg[\frac{1}{N_1}G_{ab}(X)(\dot{X}^a\dot{X}^b-2N_2\dot{X}^aX'^b+(N_2^2-N_1^2)X'^aX'^b)
+2B_{ab}\dot{X}^aX'^b\bigg], \label{f4}\ee where dot and prime
means temporal and spatial derivatives, respectively. The
canonical momenta $\pi^{\al}$ and $p_a$ conjugate to $N_{\alpha}$
and $X^a$ are:  \be \begin{array}{l} \pi^{\al}=0  \ \ \ \ \
\ \ \ \ \ \ \alpha=1,2,3 \\
p_i=\frac{1}{N_1}(2\dot{a}_i+\dot{u}\ep_{ij}a_j)-\frac{N_2}{N_1}(2a'_i+u'\ep_{ij}a_j)+2u\ep_{ij}a'_j \\
p_u=\frac{1}{N_1}(2b\dot{u}+2\dot{v}+\ep_{ij}\dot{a}_ia_j)-\frac{N_2}{N_1}(2bu'+2v'+\ep_{ij}a'_ia_j) \\
p_v=\frac{2\dot{u}}{N_1}-\frac{N_2}{N_1}2u'. \\
\end{array} \label{f8}\ee
The Canonical Hamiltonian reads: \be H=\int d^2\si
\frac{1}{N_1}G_{ab}(F^aF^b-(N_2^2-N_1^2)X'^aX'^b), \label{f6}\ee
where \be
F^a=\dot{X}^a=N_1(G^{-1})^{ab}(p_b-B_{bc}X'^c)+N_2B_{ab}X'^b
\label{f7}\ee In terms of component fields $a_i$, $u$ and $v$ we
have \be H=\int d^2\si (N_1\Psi^1+N_2\Psi^2) \label{f9}\ee where
\be \begin{array}{l}
\Psi^1={1\over4}p_i^2+{1\over4}\ep_{ij}p_va_ip_j+{1\over2}p_up_v-{1\over4}bp_v^2+{1\over16}a_i^2p_v^2 \\
\hspace{8mm}+u'\ep_{ij}a'_ia_j+\ep_{ij}ua'_ip_j+\frac{1}{2}up_va'_ia_i+(1+u^2)a'^{2}_i+bu'^2+2u'v'\\
\Psi^2=a'_ip_i+u'p_u+v'p_v, \\
\end{array} \label{f10}\ee
As can be seen from Eqs. (\ref{f8}) the momenta $\pi^{\alpha}$ are
primary constraints. The dynamics of the system is achieved by the
total Hamiltonian: \be H_T=H+\int
d\si\lambda_{\alpha}\pi^{\alpha}(\si,\ta),\label{f0110}\ee in
which $\lambda_{\alpha}$ are Lagrange multipliers. As usual we
should impose the consistency conditions on the constraints so
that they remain valid during the time. For this reason we demand
$\dot{\pi}^{\alpha}\approx 0$, where $\approx$ means weak equality
i.e. equality on the constraint surface. Using Eqs. (\ref{f0110})
and (\ref{f6}) we have: \be\begin{array}{l}
 \dot{\pi}^1=\{\pi^1,H_T\}=-\Psi^1 \\
 \dot{\pi}^2=\{\pi^2,H_T\}=-\Psi^2  \\
\dot{\pi}^3=\{\pi^3,H_T\}=0,  \\
\end{array}   \label{f0010}\ee
Therefore, the consistency of three primary constraints
$\pi^{\al}$ gives two second level constraints $\Psi^1$ and
$\Psi^2$. In this way we have so far two levels of constraints as
 \be \begin{array}{lll} \pi^1 & \pi^2 & \pi^3 \\
 \Psi^1 & \Psi^2 &\end{array}\ . \label{f0310}\ee
In order to investigate the consistency of second level
constraints, we need to calculate the Poisson brackets of
$\Psi^1(\si,\ta)$ and $\Psi^2(\si,\ta)$ at different points.
Direct calculation, using the fundamental Poisson brackets among
the four conjugate pairs $(u,p_u)$, $(v,p_v)$ and $(a_i,p_i)$
gives: \be\begin{array}{l}
 \{\Psi^1(\si,\ta),\Psi^1(\si',\ta)\}=\frac{1}{2}(\Psi^2(\si,\ta)\partial_{\si}
 -\Psi^2(\si',\ta)\partial_{\si'})\de(\si-\si') \\
 \{\Psi^1(\si,\ta),\Psi^2(\si',\ta)\}=\Psi^1(\si,\ta) \partial_{\si}\de(\si-\si') \\
\{\Psi^2(\si,\ta),\Psi^2(\si',\ta)\}=\frac{1}{2}(\Psi^2(\si,\ta)\partial_{\si}
 -\Psi^2(\si',\ta)\partial_{\si'})\de(\si-\si'),  \\
\end{array}   \label{f0015}\ee
where $\de '(\si-\si')\equiv\frac{\partial}{\partial\si}
\de(\si-\si')$. It should be noted that each of the above Poisson
brackets leads to a set of terms at different points $\si$ and
$\si'$ multiplied by $\frac{\partial}{\partial\si}\de(\si-\si')$
or $\frac{\partial}{\partial\si'}\de(\si-\si')$ which equals to
$-\frac{\partial}{\partial\si}\de(\si-\si')$. However, since these
terms have only non vanishing value when $\si'$ approaches to
$\si$, one can consider all of them at the same point. Then they
add up to give the above results. The algebra (\ref{f0015}) shows
that $\Psi^1(\si,\ta)$ and $\Psi^2(\si,\ta)$ are first class
constraints. Moreover, from (\ref{f9})  we see that:
 \be
\begin{array}{l}
 \{\Psi^1,H\}={N_{2}}'\Psi^1+{N_{1}}'\Psi^2+\frac{1}{2}N_1{\Psi^{2'}}\approx0 \\
 \{\Psi^2,H\}={N_{1}}'\Psi^1+{N_{2}}'\Psi^2+{N_{1}}\Psi^1+\frac{1}{2}N_2{\Psi^{2'}}\approx0
 \end{array} \ee
This shows that the consistency of $\Psi^1(\si,\ta)$ and
$\Psi^2(\si,\ta)$ does not give any new constraint, and we are
left with the five first class constraints given in (\ref{f0310}).

In this way we have derived three constraint chains
$\left(\begin{array}{l}
   \pi^1 \\
   \Psi^1
 \end{array}\right)$ , $\left(\begin{array}{l}
   \pi^2 \\
   \Psi^2
\end{array}\right)$ and $\left(\begin{array}{l}
   \pi^3
\end{array}\right)$ in the terminology of reference \cite{L2}.
In fact, the chain relation $\{\pi^{\alpha},H\}=\Psi^{\alpha}$
holds for all of the chains. However the first two chains are
correlated, since the Poisson bracket of the last element of each
chain with the Hamiltonian contains the other constraint. This
means that it is not possible to construct closed algebra within
each chain. The last chain contains just one element and is not
correlated to other chains, since it commutes with all of them as
well as with Hamiltonian.

As in ordinary Polyakov string one can show that $\pi^3$ generates
the Weyl symmetry of the model which affects only the components
of the world-sheet metric. In terms of the variables $N_{\al}$ we
have $N_3\rightarrow N_3+\epsilon$ under Weyl transformation. On
the
other hand the constraint chains $\left(\begin{array}{l} \pi^1 \\
\Psi^1\end{array}\right)$ , $\left(\begin{array}{l} \pi^2 \\
\Psi^2\end{array}\right)$ can be shown that generate the effect of
reparametrization invariance on the metric variables $N_1$ and
$N_2$ as well as the variables $X_a$.

\section{Gauge fixing}
We began the theory with 14 field variables in the phase space,
i.e. $X^a$, $N_{\alpha}$ and their corresponding momentum fields
$p_a$ and $\pi^{\alpha}$. Then we derived 5 first class
constraints given in (\ref{f0310}). As is well known from Dirac
theory  the first class constraints are generators of gauge
transformations \cite{HTZ1}. One needs to consider additional
conditions to fix the gauges. These "gauge fixing conditions" are
functions of phase space variables which should vanish to fix the
gauges. The gauge fixing conditions should fulfill two conditions.
First, they should constitute a system of second class constraints
when added to the original first class constraints of the system.
This condition is necessary to fix the values of variables which
vary under the action of gauge generators \cite{sh3}. Second, they
should have a closed algebra under the consistency conditions,
i.e. under the successive Poisson brackets with the Hamiltonian.

For a "complete gauge fixing" the number of independent gauge
fixing conditions should be equal to the number of first class
constraints \cite{Sh4}. In this way, we should suggest 5 gauge
fixing conditions to fix the gauges generated by the constraints
given in (\ref{f0310}), and reach a "reduced phase space" of 4
field variables. Since the momenta $\pi^{\alpha}$ are generators
of transformations in $N_{\al}$, we fix the corresponding gauge by
choosing the values of $N_{\al}$ as $N_1\approx 1,\ N_2\approx 0$
and $N_3 \approx 1$. These values are chosen such that
$g_{ij}=\eta_{ij}$. In this way we have so far introduced three
gauge fixing conditions
 \be
\begin{array}{l}
 \Omega_1\equiv N_1-1 , \\
 \Omega_2 \equiv N_2 ,\\
 \Omega_3 \equiv N_3-1.
 \end{array} \ee
It can easily seen that the system of 6 constraints $\pi^{\al}$
and $\Omega_{\al}$ are second class. The consistency of
$\Omega_{\al}$'s by the use of total Hamiltonian (\ref{f0110})
determines the lagrange multipliers $\lambda_{\alpha}$ to be zero
and does not give any new constraint. This makes us sure that the
two criterions of a good gauge mentioned above are satisfied. In
fact, by the above gauge fixing three degrees of freedom $N_{\al}$
are removed completely from the theory. This gauge has fixed the
Weyl symmetry as well as the effect of the reparametrization on
the metric variables $N_1$ and $N_2$. On the other hand, we are
still left with the remaining gauges generated by $\Psi^1$ and
$\Psi^2$ which generate the effect of reparametrization on the
variables $X_a$. In fact, since we have fixed the gauge from the
middle of the constraint chains, the gauge is fixed partially in
the language of reference \cite{Sh4}. In partial gauge fixing the
Lagrange multipliers are determined while the variations generated
by some of the gauge generators are not fixed.

To fix the effect of the parametrization of the world-sheet on
$X_a$'s, as in so many models in string theory we need to
determine some definite combinations of fields as the time
variable in target space. Taking a look on the form of the
constraints $\Psi^1$ and $\Psi^2$ in (\ref{f10}) shows that the
choice $u=\mu\ta$ is more economical in the sense that simplifies
the constraints better. Here $\mu$ is a parameter with dimension
of $(\mbox{length})^{-1}$. We recall that all of the dynamical
variables in the action are dimensionless. Hence, we consider the
gauge fixing condition
 \be \Omega_4=u-\mu\ta . \label{f0016} \ee
To fulfill the second criterion of a good gauge we choose the last
gauge fixing condition as
 \begin{eqnarray}
 \Omega_5 &\equiv &\dot{\Omega}_4 \nonumber\\ { }&=&\{\Omega_1,H_T\}
 +\frac{\partial\Omega_1}{\partial\ta} \label{f0116}\\
   {} &\approx & p_v-2\mu \nonumber
  \end{eqnarray}
This new constraint should also be valid during the time. Since
 \be \dot{\Omega}_5=2\mu(-\frac{N_2}{N_1}+N'_2)
 \approx 0 , \label{f0216} \ee
the chosen gauges are consistent and make a closed algebra with
the Hamiltonian. It is also clear that $\Omega_4$ and $\Omega_5$
make a second class system with $\Psi^1$ and $\Psi^2$. Imposing
strongly the constraints (\ref{f0016}) and (\ref{f0116}) on the
system, simplifies the constraints $\Psi_1$ and $\Psi_2$ as
  \be   \begin{array}{l}
 \Psi^1\rightarrow\bar{\Psi}^1=\frac{1}{4}p_i^2+\frac{1}{2}\ep_{ij}
 \mu a_ip_j+\ep_{ij}\mu\ta a'_ip_j+(1+\mu^2\tau^2)a'^{2}_i+\mu p_u-b\mu^2+\frac{1}{2}\mu^2a_i^2+\mu^2\tau a_ia'_i, \\
 \Psi^2\rightarrow\bar{\Psi}^2=a'_ip_i+2\mu v',
 \end{array} \label{f18}\ee
This shows that $p_u$ and $v$ can be derived on the constraint
surface, i.e. from identities $\bar{\Psi}_1=0$ and
$\bar{\Psi}_2=0$, in terms of the physical variables $a_i$ and
$p_i$. In this way the reduced phase space is just the four
dimensional space of $(a_i,p_i)$ whose original Poisson brackets
serve as the Dirac brackets in the remaining physical space. The
terms $\mu p_u$ and $\mu^2 b$ in the expressions of $\bar{\Psi}_1$
have nothing to do with the dynamics of $(a_i,p_i)$  and can be
dropped. The parameter $b$ has in fact no important role in the
theory and only shifts the spectrum of the energy with a constant
value.

As in reference \cite{ref1} we consider the dimensionless quantity
$\mu l$ as a small parameter which should be considered only in
the first order. Therefore, in all of the foregoing calculations
we keep only linear terms with respect to $\mu$, assuming that $l$
is finite. Therefore, the Hamiltonian (\ref{f9}) in the reduced
phase space can be written in terms of the Hamiltonian density:
 \be
 \mathcal{H}_{GF}=\frac{1}{4}p_i^2+\frac{1}{2}\ep_{ij}\mu
 a_ip_j+\ep_{ij}\mu\ta a'_ip_j+a'^{2}_i. \label{f18}\ee
Since $B(X)$ in (\ref{f2}) is linear with respect to $u$ one may
think of $\mu$ as the order of magnitude of the $B$-field. This
assumption is equivalent to considering the effect of the
$B$-field only up to the first order.
\section{Boundary conditions as constraints \label{sec6}}
From now on we forget about the original theory and suppose we are
given a theory with two degrees of freedom $a_i$ and the
corresponding momenta $p_i$ whose dynamics is given by
the final Hamiltonian (\ref{f18}). We make a change of
variables from $(a_i,p_i)$ to $(A_i=\ep_{ij}a_j,P_i=p_i)$. Then the the
fundamental Poisson brackets which is the same as the final Dirac
bracket of the original theory read
\be \begin{array}{l}  \{A_i(\si,\tau), P_j(\si',\tau)\}=\ep_{ij}\de(\si-\si'),\\
   \{A_i(\si,\tau),A_j(\si',\tau)\}=\{P_i(\si,\tau),P_j(\si',\tau)\}=0  \end{array} \label{f22a}\ee
The Hamiltonian equation of motion for the remaining fields, can
be written as
 \be
\begin{array}{l}
\dot{A}_i={1\over 2}\ep_{ij}(P_j-2\mu\ta A'_j-\mu A_j) \\
\dot{P}_i=-\ep_{ij}({1\over 2}\mu P_j-\mu\ta P'_j+2A''_j) \\
\end{array}  \label{f23}\ee

The only things that should be brought from the original theory
are the boundary conditions. Using the original action (\ref{f4})
the boundary condition after gauge fixing emerge in terms of phase
space variables as:
 \be \Phi_i^{(1)}=\mu\ta
 P_i-2A'_i=0 \;\;\;\;\;\ \mbox{at $\si  =0,l$} \label{f24}\ee
We have shown in the appendix that the boundary condition
(\ref{f24}) can also be derived from the parallel approach as the
equations of motion of the end points in the discretized version.

As mentioned in the introduction we do not want to find the
general solution of the dynamical equations of motion. On the
other hand, we are interested to follow the dynamics of the
boundary conditions which means investigating the consistency of
primary constraints $\Phi^{(1)}_i (\sigma)|_{\sigma=0}$ and
$\Phi^{(1)}_i (\sigma)|_{\sigma=l}$. Using the gauge fixed
Hamiltonian of the previous section (\ref{f18}) the total
Hamiltonian at this stage is
 \be \overline{H}_T= \int_0^l d\sigma[\frac{1}{4}P_iP_i-
 \frac{1}{2}\mu A_iP_i-\mu\ta A'_iP_i
 +A'_iA'_i]+\Lambda_1^i \Phi^{(1)}_i (\sigma)|_{\sigma=0}
 +\Lambda_2^i \Phi^{(1)}_i (\sigma)|_{\sigma=l} .\label{s1} \ee
The consistency of primary constraints for instance at $\si=0$
gives
  \be 0= \left[ \mu P_i-\ep_{ij}P'_j+\mu\ep_{ij}A'_j \right]_{\si=0}
  +\Lambda_1^j\  \left\{\Phi^{(1)}_i|_{\si=0}\ ,\Phi^{(1)}_j |_{\si=0}
  \right\} \label{f0024}\ee
Similar equations should be written at the end-point $\si=l$. As
discussed in details in \cite{ref4} the first term in the LHS of
Eq. (\ref{f0024}) has not the same order as the coefficient of
$\Lambda^i_1$ (and $\Lambda^i_2$) in the second term when
regularizing the Dirac delta function. Therefore this condition
can be fulfilled identically only if $\Lambda^i_{1,2}$ as well as
the first term vanish simultaneously. In this way we have used the
consistency conditions of the constraints for simultaneously
determining the undetermined Lagrange multiplier and finding the
next level of constraints as $\Phi_i^{(2)}(0)$ and
$\Phi_i^{(2)}(l)$ where
 \be \Phi_i^{(2)}(\sigma) =P_i-\ep_{ij}P'_j+\mu\ep_{ij}A'_j .\ee
Then we should consider the consistency of second level
constraints by using the Hamiltonian
 \be \overline{H}= \int_0^l d\sigma[\frac{1}{4}P_iP_i-
 \frac{1}{2}\mu A_iP_i-\mu\ta A'_iP_i
 +A'_iA'_i] \label{s2} \ee
which is the same as the total Hamiltonian (\ref{s1}) after
imposing $\Lambda^i_{1,2}=0$. This gives the third level of
constraints. Subsequent levels of constraints can be derived in
the same way. Using the relations:
  \be \begin{array}{l}\{A_i^{(n)},\overline{H}\}=\frac{1}{2}
  \ep_{ij}(P^{(n)}_j-\mu A_j^{(n)}-2\mu\ta A^{(n+1)}_j)+{\cal O}(\mu^2) \\
  \{P_i^{(n)},\overline{H}\}=-\ep_{ij}(\frac{1}{2}\mu P^{(n)}_j-\mu\ta
  P^{(n+1)}_j+2A^{(n+2)}_j)+{\cal O}(\mu^2), \\ \end{array}
  \label{f25}\ee
where $A_i^{(n)}=\partial_{\sigma}^nA_i$ and
$P_i^{(n)}=\partial_{\sigma}^nP_i$ one can inductively show that
the full set of constraints are $\Phi_i^{(N)}(0)\approx0$ and
$\Phi_i^{(N)}(l)\approx0$ where
  \be \begin{array}{ll}
  \Phi_i^{(2n+1)}=-n\mu P_i^{(2n-1)}+ \mu\ta P_i^{(2n)}-2n\mu
  \ep_{ij}A^{(2n)}_j-2A^{(2n+1)}_i+{\cal O}(\mu^2), &  \\
  \Phi_i^{(2n+2)}=(n+1)\mu P^{(2n)}_i-\ep_{ij}P^{(2n+1)}_j
  +(2n+1)\mu\ep_{ij}A_j^{(2n+1)}+{\cal O}(\mu^2) & n=0,1,2,
  \cdots \\ \end{array} \label{f26}\ee
For practical calculations we write the constraints as ordinary
functions in the bulk of the string and then integrate them with
the use of $\de(\si)$ and $\de(\si-l)$ respectively.

Now we want to investigate whether the constraints are first or
second class. For this reason one should calculate the Poisson
brackets of the constraints. Since the constraints contain
different orders of derivatives of $A_i(\si,\tau)$ and
$P_i(\si,\tau)$, the Poisson brackets
$C_{ij}^{k,k'}\equiv\{\Phi_i^{k},\Phi_j^{k'}\}$ contain
derivatives of orders $k+k'$, $k+k'-1$, etc, of the Dirac delta
function, which are highly divergent and independent of each
other. One way of treating the matrix of Poisson brackets is
regularizing the delta functions as gaussian functions of width
$\varepsilon$ and let $\varepsilon\rightarrow 0$ after all. A
tedious calculation gives
 \be \begin{array}{l}
 C^{2m+1,2n+1}_{ij}=\frac{-2\mu\ep_{ij}}{\sqrt{\pi}}\varepsilon^{-2(m+n+1)}
 (\varepsilon (m+n)H_{2m+2n}(0)-2\tau H_{2m+2n+1}(0))+ \mathcal{O}(\mu^2) \\
  C^{2m+2,2n+1}_{ij}=\frac{-2}{\sqrt{\pi}}\varepsilon^{-2(m+n+1)-1}
 (n\mu\varepsilon\ep_{ij} H_{2m+2n+1}(0)+\de_{ij} H_{2m+2n+2}(0))+ \mathcal{O}(\mu^2),\\
 C^{2m+2,2n+2}_{ij}=\frac{2\mu\ep_{ij}}{\sqrt{\pi}}\varepsilon^{-2(m+n+1)-1}
  H_{2m+2n+2}(0)+ \mathcal{O}(\mu^2)
 \end{array}  \label{f27}\ee
where $H_n(x)$ are Hermite polynomials. Similar expressions should
be considered with $H_n(1)$ at the end-point $\si=l$. The non
vanishing elements on each row are located such that no vanishing
linear combination of rows may be found. This means that the
infinite dimensional matrix $C_{ij}^{k,k'}$ is not singular and
can in principle be inverted. Therefore, all of the constraints
are second class. However, it is not practically possible to find
the inverse of $C_{ij}^{k,k'}$. The problem is how we can find the
Dirac brackets of the fields which need to have $C^{-1}$.
\section{Reduced phase space}
As stated before, we seek for appropriate coordinates in which
imposing the constraints (\ref{f26}) lead to omitting a set of
canonical pairs. Here we have a difficult problem in which the
ordinary Fourier expansion does not do this job. However, in the
limit $\mu\rightarrow0$ the boundary condition (\ref{f24}) is the
ordinary  Neumann  one and the Hamiltonian (\ref{s2}) has a simple
quadratic form in terms of coordinates and momenta. Hence, we need
to write extended Fourier transformations for the fields $A_i$ and
$P_i$ that include at most linear corrections with respect to the
parameter $\mu$ and go to the ordinary Fourier transformation in
the limit $\mu\rightarrow0$. Since $\mu \tau$ and $\mu \sigma$ are
the only dimensionless quantities that can be used for this
correction, what can we do is correcting the Fourier coefficients
by correction terms linear in $\tau$ or $\sigma$. The linear term
in $\tau$, however, is not needed at this stage, since it can be
considered as part of the solution of the equations of motion.
Adding all these points up together we suggest the following
extended Fourier transformations for the fields
  \be   A_i(\si,\ta)=\frac{1}{\sqrt{2\pi}}\int_{-\infty}^{\infty} d k
  \left[\left(A_i(k,\ta)+\mu \si\al_i(k,\ta)\right)\cos k\si+
  \left(B_i(k,\ta)+\mu\si\beta_i(k,\tau)\right)\sin k\si\right],
  \label{f32} \ee \be P_i(\si,\ta)=\frac{-\ep_{ij}}{\sqrt{2\pi}}
  \int_{-\infty}^{\infty} d k\left[\left(C_j(k,\ta)+\mu\si\ga_j(k,\ta)\right)
  \cos k\si+\left(D_j(k,\ta)+\mu\si\de_j(k,\tau)\right)\sin k\si
  \right] .\label{f'32}\ee
In ordinary Fourier expansions the coefficients $A_i(k,\tau)$,
$B_i(k,\tau)$, $C_i(k,\tau)$ and $D_i(k,\tau)$ contain the same
amount of data as the original fields $A_i(\si,\tau)$ and
$P_i(\si,\tau)$. Comparing the expansions (\ref{f32}) and
(\ref{f'32}) with ordinary Fourier expansions shows that we have
used a duplicated basis including $\sin$'s, $\cos$'s, $\sigma$
times $\sin$'s and $\si$ times $\cos$'s for expanding our fields.
This basis is complete but its elements are not independent.
Mathematically it is allowed to use a basis which is "larger than
necessary". However, the essential point is that  one should
assume appropriate Poisson brackets among the extended Fourier
modes such that the desired fundamental Poisson brackets
(\ref{f22a}) remain valid. In other words, we should tune their
brackets in such a way that our physical phase space variables,
which are half of the extended phase space variables, do obey the
right Poisson brackets. Direct calculation shows that the
following Poisson brackets lead to the standard Poisson algebra
(\ref{f22a}) for the physical fields,
 \be \begin{array}{l}
 \{A_i(k,\ta),C_j(k',\ta)\}=\{B_i(k,\ta),D_j(k',\ta)\}=\de_{ij}\de(k-k'), \\
 \{\al_i(k,\ta),D_j(k',\ta)\}=\{\ga_i(k,\ta),B_j(k',\ta)\}=
 \de_{ij}\pa_{k'}\de(k-k'). \end{array} \label{f36}\ee
All other Poisson brackets are assumed to vanish. Specially the
modes $\beta_i$ and $\delta_i$ have vanishing Poisson brackets
with all other variables in the extended Fourier space and so
decouple from the theory. This means that we can put them away and
write down the expansions only with linear terms in the cosine
modes. We will see on the other hand that omitting the modes
$\beta_i$ and $\delta_i$ does not disturb our analysis of imposing
the boundary conditions. We have, up to this point, 6 sets of real
variables in the extended Fourier space which depend on real,
continues and positive variable $k$.

Now we want to impose the full set of constraints (\ref{f26}) on
the fields. Using the expansions (\ref{f32}) and (\ref{f'32}) the
constraints at the end-point $\si=0$ lead to
 \be \begin{array}{l}\int_{-\infty}^{\infty} d k k^{2n}\left[\mu\ta\ep_{ij}C_j+
 2n\ep_{ij}A_j+(4n+2)\al_i+2k\tilde{B}_i\right]+{\cal O}(\mu^2)=0
 \\
 \int_{-\infty}^{\infty} d k k^{2n-1}\left[(n+1)\ep_{ij}C_j+(2n+1)\ga_i
 +k\tilde{D}_i\right]+{\cal O}(\mu^2)=0
 \end{array} \label{f0036}\ee
where $B_i=\mu \tilde{B}_i$ and $D_i=\mu\tilde{D}_i$. Since these
conditions should be satisfied for arbitrary values of $n$ we have
 \be \begin{array}{l}
 \mu\ta\ep_{ij}C_j+2n\ep_{ij}A_j+(4n+2)\al_i+2k\tilde{B}_i=0,  \\
 (n+1)\ep_{ij}C_j+(2n+1)\mu\ga_i+k\tilde{D}_i=0.
  \end{array} \label{f37}\ee
The difficulty arises here since the integer $n$, which shows the
level of constraints, has appeared in the form of relations among
the Fourier modes. This means that it is not possible to satisfy
the constraints of all levels just by considering simple linear
relations among the Fourier modes of a given $k$ as can be done in
ordinary Dirichlet,  Neumann, or even mixed boundary conditions
\cite{ref3}. In fact, this phenomenon is the reason which makes
the ordinary Fourier expansion inadequate for realizing the
constraints. However, we have the opportunity of existence of
extra variables in the extended phase space, which provides us
additional tools for satisfying the constraints. In this way we
are allowed to assume that the coefficients of $n$ besides the
terms independent of $n$ in (\ref{f37}) vanish. This gives
 \be \begin{array}{ll}\al_i=-\frac{1}{2}\ep_{ij}A_j+
 {\cal O}(\mu^2) & \tilde{B}_i=\frac{1}{2k}\ep_{ij}
 (A_j-\ta C_j)+{\cal O}(\mu^2) \\ \ga_i=-\frac{1}{2}\ep_{ij}C_j
 +{\cal O}(\mu^2) & \tilde{D}_i=-\frac{1}{2k}\ep_{ij}C_j
 +{\cal O}(\mu^2)   \end{array}   \label{f38}\ee
Hence the main fields $A_i(\si,\tau)$ and $P_i(\si,\tau)$ can be
written in terms of two remaining sets of Fourier modes
$A_i(k,\tau)$ and $C_i(k,\tau)$ as
 \be   A_i(\si,\ta)=\frac{1}{\sqrt{2\pi}}\int_{-\infty}^{\infty} d k
  \left[(\delta_{ij}-\frac{1}{2}\mu\si\epsilon_{ij})A_j\cos k\si+
  \frac{\mu}{2k}\epsilon_{ij}(A_j-\tau C_j)\sin k\si\right],
  \label{s32} \ee \be P_i(\si,\ta)=\frac{-1}{\sqrt{2\pi}}
  \int_{-\infty}^{\infty} d k\left[ (\ep_{ij}+\frac{1}{2}\mu\si\de_{ij})C_j
  \cos k\si+\frac{\mu}{2k}C_i\sin k\si
  \right] .\label{s'32}\ee
As expected, the zeroth order (with respect to $\mu$) of the Eqs.
(\ref{s32}) and (\ref{s'32}) is the expansion of a simple bosonic
string with Neumann boundary condition at the end point $\si =0$.
The linear term with respect to $\si$ in cosine modes as well as
the sin term itself are appeared as the first order corrections.

Next we should impose the constraints (\ref{f26}) at the end-point
$\si=l$ on the fields derived recently in Eqs. (\ref{s32}) and
(\ref{s'32}). Hence we find
 \be \begin{array}{l} \int_{-\infty}^{\infty} d k
 k^{2n-1}(-1)^n[n\mu\ep_{ij}C_j+2k^2(A_i-\frac{1}{2}\mu\si A_j)]
 \sin (kl)+{\cal O}(\mu^2) =0, \\  \\ \int_{-\infty}^{\infty} d k
 k^{2n+1}(-1)^n[(\de_{ij}-\frac{1}{2}\mu\si\ep_{ij})
 C_j-(2n+1)\mu\ep_{ij}A_j] \sin (kl)+{\cal O}(\mu^2)=0.
 \end{array} \label{f0038}\ee
The above constraints are satisfied identically for $kl=m\pi$.
However, for $k\neq \frac{m\pi}{l}$ there is no way for satisfying
the constraints for arbitrary $n$ except assuming that
 \be A_i(k, \tau)= C_i(k, \tau)=0 \hspace{1cm} \mbox{for}\ \ \
 k\neq \frac{m\pi}{l} \label{s4} \ee
This leads to descritizing the Fourier modes.

Before writing the final form of the fields in terms of the set of
enumerable Fourier modes, care is needed to write the zero modes.
The contributions due to cosine modes come out automatically by
letting $k=0$. However, contributions to zero modes originating
from sine terms should be derived by taking the following limits:
 \be \lim_{k\rightarrow 0}\tilde{B}_i\sin k\si=
 \frac{1}{2}\si\ep_{ij}(A_j(0, \tau)-\ta C_j(0,\tau)),
 \hspace{1cm} \lim_{k\rightarrow 0}
 \tilde{D}_i\sin k\si=- \frac{1}{2}\si\ep_{ij}C_j(0,\tau),
    \label{f39}\ee
which follow from Eqs.(\ref{f38}). Adding these two contributions
the zero mode part of the fields are so far as follows
 \be \begin{array}{l} A^0_i(\si,\ta)= A_i^0(\ta)-\frac{1}{2}\mu\si\ta
 \epsilon_{ij}C^0_j(\ta)\\ P^0_i(\si,\ta)=-(\ep_{ij}+\mu\si\de_{ij})C^0_j(\ta) \end{array} \label{s10} \ee

At this point we want to notice the reader to a global symmetry of
the gauged fixed Lagrangian. If we turn off the B-field we would
have an ordinary bosonic string in which only the derivatives of
the A-fields are present in the Lagrangian. This allows one to
shift the fields by a constant amount without any change in the
Lagrangian. When the B-field is on, Eq. (\ref{f18}) shows that the
A-field itself is present in the gauged fixed Hamiltonian.
However, the relevant term, i.e. the second term in Eq.
(\ref{f18}), is proportional to $\mu$. This shows that the theory
is symmetric, up to second order terms with respect to $\mu$,
under the following transformation
 \be A_i(\si,\ta) \rightarrow A_i(\si,\ta)+ \mu f(\ta) \label{s11}
 \ee
where $f(\ta)$ is an arbitrary function of time. This symmetry
leads to an ambiguity in the zero mode of the A-field. Hence we
should correct the first row of Eq. (\ref{s10}) in the most
general case as follows
 \be A^0_i(\si,\ta)= A_i^0(\ta)-\frac{1}{2}\mu\si\ta
 \epsilon_{ij}C^0_j(\ta) + \mu l [(a_{ij} A^0_j(\ta) +
 b_{ij} C^0_j(\ta)] \ee
Note that $\mu l$ is the only relevant dimensionless quantity
which is first order in $\mu$. The unknown coefficients $a_{ij}$
and $b_{ij}$ should be determined upon suitable assumptions about
the algebra of the fields. The best assumption seems to be keeping
the standard algebra (\ref{f22a}) in the bulk of the string and
letting all changes in the algebra of the fields lay on the
boundaries. If we make this choice the final form of the physical
fields in terms of the set of discrete Fourier modes
$A_i^m(\tau)\equiv A_i(\frac{m\pi}{l},\tau)$ and
$C_i^m(\tau)\equiv C_i(\frac{m\pi}{l},\tau)$ are as follows
 \be \begin{array}{lll}
 A_i(\si,\ta)&=&\frac{1}{\sqrt{l}}\bigg[A_i^0(\ta)-\frac{1}{2}
 \mu\ta(\si-\frac{l}{2})\ep_{ij}C_j^0(\ta)-\frac{1}{2}
 \mu l\ep_{ij}A_j^0(\ta)\bigg]  \\
 {}&+& \sqrt{\frac{2}{l}}\sum_{m=1}^{\infty}\bigg[(A_i^m(\tau)
 -\frac{1}{2}\mu\si\ep_{ij}A_j^m(\ta))\cos \frac{m\pi\si}{l}+
 \frac{\mu l}{2m\pi}\ep_{ij}(A_j^m(\ta)-\ta C_j^m(\ta))\sin
 \frac{m\pi\si}{l}\bigg] \end{array} \label{f41}\ee
 \be \begin{array}{lll}
 P_i(\si,\ta)&=&-\frac{1}{\sqrt{l}}\bigg[\ep_{ij}C_j^0(\ta)+\mu\si
 C_i^0(\tau)\bigg]  \\ {} &-&
 \sqrt{\frac{2}{l}}\sum_{m=1}^{\infty}\bigg[(\ep_{ij}C_j^m(\ta)
 +\frac{1}{2}\mu\si C_i^m(\ta))\cos \frac{m\pi\si}{l}+
 \frac{\mu l}{2n\pi}C_i^m(\ta)\sin \frac{m\pi\si}{l}\bigg]
 \end{array}  \label{f42}\ee
The normalization factor $\frac{1}{\sqrt{2\pi}}$ is replaced by
$\sqrt{\frac{2}{l}}$ for oscillatory modes and
$\frac{1}{\sqrt{l}}$ for zero mode upon going from the continues
parameter $k$ to the discrete parameter $m$. \footnote{Since
another length scale, i.e. $\mu^{-1}$, exists in the model, one
may suppose that the normalization factors should differ from the
ordinary Fourier series. However, it can be shown that such
corrections only changes the observables  by amounts of ${\cal
O}(\mu^2)$ which is not important} With this normalization the
brackets of the discrete modes should also be given in terms of
Kronecker delta as
 \be \{A_i^m,C_j^{m'}\}=\delta_{ij}\delta_{mm'}, \label{s5} \ee
 \be\{A_i^m,A_j^{m'}\}=\{C_i^m,C_j^{m'}\}=0. \label{s6} \ee
In fact, the remaining canonical pairs $A_i^m$ and $C_i^m$ as a
small part of the original phase space are natural coordinates of
the reduced phase space. On the other hand, a great part of the
initial phase space variables are omitted due to the constraints.

Remember that if one is able to omit the redundant variables due
to all kinds of constraints and write down the relevant fields in
terms of final canonical coordinates of the reduced phase space,
then there is no need to find the Dirac brackets. In other words,
we pay the expense of using the Dirac brackets whenever it is not
possible to find  a canonical basis to describe the reduced phase
space. Hence, we will find the Dirac brackets of the original
fields $A_i(\si,\tau)$ and $P_i(\si, \tau)$ if we calculate their
brackets  by using the brackets (\ref{s5}) and (\ref{s6}).

Eq. (\ref{f42}) shows that the momentum-fields $P_i(\si,\tau)$
just include the variables $C_i^m$ and have vanishing brackets:
 \be \{P_i(\si,\ta),P_j(\si',\ta) \}=0. \label{f43}\ee
Straightforward calculations gives the brackets of coordinate and
momentum fields as
 \be \{A_i(\si,\ta),P_j(\si',\ta)\}=\ep_{ij}\de_N(\si,\si'), \label{f44}\ee
where
 $$\de_N(\si,\si')\equiv\de(\si-\si')+\de(\si+\si'). $$
Since both $\si$ and $\si'$ lie in the interval $[0,l]$ their sum
never vanishes. So the second delta function does not have any
role and Eq. (\ref{f44}) reduces to the usual form of Eq.
(\ref{f22a}). However, since in the expansion of $A$-fields both
variables $A_i^m$ and $C_j^m$ are present, the interesting
phenomenon appears in the bracket of coordinate fields at
different points. Direct calculation gives
 \be \{A_i(\si, \ta),A_j(\si', \ta)\}=\frac{1}{2}\mu\ta\ep_{ij}
 \left(\frac{\si+\si'}{l}-1+\frac{2}{\pi}
 \sum_{n=1}^{\infty}\frac{1}{n}\sin\frac{n\pi}{l}(\si+\si') \right). \label{s7}  \ee
This result is similar to what derived in \cite{ref3} for a string
coupled to constant background B-field. The right hand side of Eq.
(\ref{s7}) vanishes in the bulk of the string, i.e. when $\si$ or
$\si'$ does not lie on the end points. It gives (-2) when
$\si=\si'=0$ and (+2) when $\si=\si'=l$. However, as the B-field
itself, the amount of non commutativity grows linearly with time.
Our result here defers from reference \cite{ref2} with a term
proportional to $\mu\tau^2$ which is the same on both boundaries
as well as in the bulk of the string. If, however, we add a term
$-\frac{1}{2} \mu\ta^2\ep_{ij}C_j^0(\ta)$ to the zero mode part of
the field $A_i(\si,\ta)$ in Eq. (\ref{f41}), our result will
coincide with reference \cite{ref2}. This correction is allowed
according to the global symmetry of Eq. (\ref{s11}). This means
that we have forgiven our previous assumption that the components
of the A-field commute in the bulk of the string. With this
assumption the resulted brackets can be summarized as follows
 \be \begin{array}{l}
 \{A_i(\si,\ta),P_j(\si',\ta)\}=\ep_{ij}\de\si,\si'),\\
 \{P_i(\si,\ta),P_j(\si',\ta) \}=0\\
 \{A_i(\si, \ta),A_j(\si', \ta)\}= \left\{ \begin{array}{ll}
  \frac{\mu\tau^2\epsilon_{ij}}{2l}&\si\neq0,l \ \ \mbox{or}\ \
  \si'\neq0,l\\ \mu\tau\ep_{ij}(1+\frac{\tau}{2l})
  &\si=\si'=0\\ \mu\tau\ep_{ij}(-1+\frac{\tau}{2l})
  &\si=\si'=l  \end{array}  \right. \end{array} \ee
This shows that the fundamental characters of the $A$-fields and
$P$-fields as coordinate and momentum fields are remained almost
as before and the time dependent B-field leads to a time dependent
non commutativity in the coordinate fields all over the string.
 \section{Concluding remarks}
In this paper we gave a complete Hamiltonian treatment of the
Nappi-Witten model (WZW model based on non semi simple gauge
group) as an interesting and non trivial system in which
complicated boundary conditions make the physical subset of
variables far from reaching. The initial dynamical variables in
this model are 4 components of a bosonic string,
$X_a=(a_1,a_2,u,v)$, and the components of world-sheet metric. We
used appropriate variables to find 3 primary and 2 secondary first
class constraints. It can be shown that these constraints are
generators of reparametrizations as well as Weyl transformations.
Then we fixed the gauge such that the world-sheet metric is flat
and $u=\mu \tau$ where the small parameter $\mu$ is proportional
to the strength of the B-field. In this way the components of the
world-sheet metric and the variables $u$ and $v$ disappeared as
the result of constraints and gauge fixing conditions. Hence, we
derived a smaller theory with two coordinate fields $a_1$ and
$a_2$ and their corresponding momentum fields.

The most important part of the problem seems to be the boundary
conditions which should be brought from the original theory.
Considering the boundary condition as Dirac constraints and
following their consistency, we found two infinite chains of
second class constraints at the end-points which restricted the
space of physical variables to a much smaller set. Due to
complicated form of the boundary conditions, it is not an easy
task to impose them on the space of the physical variables. In
fact, with an ordinary Fourier expansion the constraints do not
lead simply to omitting some Fourier modes as in Dirichlet or
Neumann boundary conditions.

To overcome this difficulty we extended the phase space to a
larger one which is given by an extended Fourier expansion in
which the Fourier modes are replaced by linear functions of the
variables. In this basis the infinite set of constraints can be
imposed more easily by using the arbitrariness due to extra
variables. This results to disappearing of so many canonical pairs
among the used extended Fourier basis and finally a set of
discrete modes remain which act as the canonical coordinates of
the reduced phase space. Then all physical objects including the
original coordinate and momentum fields can be expanded in terms
of these modes.

Using these expansions we found that the commutation relations of
the coordinate and momentum fields are almost as usual, except
that the coordinate fields do not commute at the boundaries, with
an amount proportional to time and/or B-field but with opposite
signs at two boundaries. We showed that it is allowed to insert a
term which gives non commutativity proportional to $\tau^2$
throughout the string. This correction may make our results
consistent with those of reference \cite{ref2} in which the
authors have given iterative solutions for the equations of
motion.

We think that our method here has two main advantages in two
different areas. First, we do not solve the equation of motion.
Therefore, in our final result the time dependence of remaining
modes are not specified. However, this time dependence is not
essential for quantization of the model. If needed, one can use
the Hamiltonian written in terms of the final modes and then
derive their time dependence. In fact, our main objective is that
for quantizing a theory, i.e. investigating the algebraic
structure of the observables, it is not needed to follow the full
dynamics of the system; it is just enough to study the dynamics of
constraints. As a matter of fact, for simple models it may seem
more simple and economic to solve the equations of motion and then
quantize the theory, since this procedure contains the dynamics of
the constraints within itself. But this may not be the case for a
complicated model such as the model considered in this paper.

The next advantage is in the context of constraint systems. As we
see in the literature \cite{ref0,ref4} the main difficulty in
considering the infinite set of constraints due to boundary
conditions is deriving the Dirac brackets. In this paper, as in
our previous work \cite{ref3} we showed that if one is able to
find a set of canonical variables describing the reduced phase
space, then there is naturally no need to calculate the Dirac
brackets. In fact, this was the main brilliant idea of Dirac
\cite{Dirac}, who gave his famous formula of Dirac brackets in
such a way that it is equivalent to calculating the Poisson
brackets only in the space of canonical variables describing the
reduced phase space.

 \end{document}